\documentclass[conference]{IEEEtran}
\IEEEoverridecommandlockouts
\usepackage{cite}
\usepackage{amsmath,amssymb,amsfonts}
\usepackage{algorithmic}
\usepackage{textcomp}
\usepackage{xcolor}

\usepackage[T1]{fontenc}

\usepackage{amsmath,amsfonts}
\usepackage{algorithmic}
\usepackage{array}
\usepackage[caption=false,font=normalsize,labelfont=sf,textfont=sf]{subfig}
\usepackage{booktabs}
\usepackage{textcomp}
\usepackage{stfloats}
\usepackage{multirow}
\usepackage[ruled,linesnumbered]{algorithm2e}
\usepackage{verbatim}
\usepackage{graphicx}
\usepackage{subfig} 
\usepackage{subfloat}
\usepackage{float}
\usepackage{bm}
\def\BibTeX{{\rm B\kern-.05em{\sc i\kern-.025em b}\kern-.08em
    T\kern-.1667em\lower.7ex\hbox{E}\kern-.125emX}}
\begin{document}

\title{Contrastive Latent Space Reconstruction Learning for Audio-Text Retrieval\\
}

\author{\IEEEauthorblockN{Kaiyi Luo$^{1,2\dagger}$, Xulong Zhang$^{1\dagger}$\thanks{$^\dagger$ Both authors have made equal contributions.}, Jianzong Wang$^{1\ast}$\thanks{$^{\ast}$Corresponding author: Jianzong Wang (jzwang@188.com)}, Huaxiong Li$^{2}$, Ning Cheng$^{1}$, Jing Xiao$^{1}$}
	\IEEEauthorblockA{\textit{$^{1}$Ping An Technology (Shenzhen) Co., Ltd. } \\ 
		\textit{$^{2}$Department of Control Science and Intelligent Engineering, Nanjing University}
	}
}

\maketitle

\begin{abstract}
Cross-modal retrieval (CMR) has been extensively applied in various domains, such as multimedia search engines and recommendation systems. Most existing CMR methods focus on image-to-text retrieval, whereas audio-to-text retrieval, a less explored domain, has posed a great challenge due to the difficulty to uncover discriminative features from audio clips and texts. Existing studies are restricted in the following two ways: 1) Most researchers utilize contrastive learning to construct a common subspace where similarities among data can be measured. However, they considers only cross-modal transformation, neglecting the intra-modal separability. Besides, the temperature parameter is not adaptively adjusted along with semantic guidance, which degrades the performance. 2) These methods do not take latent representation reconstruction into account, which is essential for semantic alignment. This paper introduces a novel audio-text oriented CMR approach, termed Contrastive Latent Space Reconstruction Learning (CLSR). CLSR improves contrastive representation learning by taking intra-modal separability into account and adopting an adaptive temperature control strategy. Moreover, the latent representation reconstruction modules are embedded into the CMR framework, which improves modal interaction. Experiments in comparison with some state-of-the-art methods on two audio-text datasets have validated the superiority of CLSR.
\end{abstract}


\section{Introduction}
The explosion of multi-modal data has sparked significant interest in cross-modal retrieval, which is aimed to conduct similarity searches across different modalities \cite{cou1,cou2,cou3}. For example, in a search engine, one expects relevant results such as videos and pictures with text inputs. However, the heterogeneous data structures of different modalities pose a great challenge to similarity measurement \cite{cou4}. The fundamental goal of cross-modal retrieval is to establish correlations between diverse forms of data, such as images, texts, audios, and videos. A frequently used approach involves creating a shared latent space in which instances with similar semantics from various modalities are positioned close to one another, enabling the system to perform efficient retrieval tasks \cite{cou5}. Given a textual query, a cross-modal retrieval system can transform the textual feature into a common space, and decide its corresponding image by computing distances among data.

In recent years, numerous CMR methods have been proposed \cite{AGCH,MCCN,con1,LCFS,AMSH}, which fall into two categories: optimization-based methods and deep methods. optimization-based methods such as \cite{LCFS} reframe the challenge of CMR as an optimization task, which can be solved efficiently with different optimization schemes such as Alternating Direction Method of Multipliers (ADMM). Deep learning based methods extract robust representations with neural networks, which contains less noise and more discriminative features. Generally, deep methods outperform shallow ones due to the powerful representation capacity of neural networks. Though the existing CMR methods can achieve satisfying performance, these studies are focused on text-image retrieval and text-video retrieval, without considerations for audio-text retrieval, which is more challenging because audio information may contain noise and multiple sources. To our knowledge, there are several works dedicated to audio-text retrieval \cite{related1,related2,related3,related4}. The basic method is to apply different network structures and contrastive learning to unify features from the two modalities. Despite the good performance produced by these methods, there are several problems to be solved: 1) Most researchers utilize contrastive learning to construct a common subspace where similarities among data can be measured. However, they only considers cross-modal transformation, neglecting the intra-modal separability. Besides, the temperature parameter is not adaptively adjusted along with semantic guidance, which degrades the performance. 2) These methods do not take latent representation reconstruction into account, which impairs the performance of these methods.

To tackle the aforementioned challenges, this paper introduces a novel approach for audio-text retrieval, termed as Contrastive Latent Space Reconstruction (CLSR). For data processing, raw audio data are transformed into log Mel-spectrograms, which depict fine-grained acoustic features and can be processed like images. textual data are converted into word embeddings with the BERT model \cite{bert}, which contain high-level correlations between words. To construct the shared space, CLSR extends the existing NT-Xent contrastive loss\cite{XNENT} to increase feature discrimination, and adopts adaptive temperature control, which increases the positive compactness and negative separability. Moreover, a latent reconstruction module is designed for each modality for better semantic alignment. The main contribution of CLSR can be summarized as follows:
\begin{itemize}
	\item We introduce a novel CMR method, i.e., CLSR, which provides a new perspective for audio-text CMR problems.
	\item Considering the property of audio-text datasets, we adopt a temperature-adaptive contrastive method, which enhances the discrimination of latent representations based on semantic alignment.
	\item Experimental results across multiple datasets illustrate that CLSR surpasses certain state-of-the-art cross-modal methods.
\end{itemize} 
\section{Related Work}
\subsection{Audio-Text Retrieval}
Despite the effectiveness of the previous methods, audio-text retrieval is less studied due to the difficulty of discovering discriminative representations for audio clips. Chechik et al. \cite{related1} proposed a scalable machine learning method with the passive-aggressive model (PAMIR). Ikawa et al. \cite{related2} constructs a latent variable space with the onomatopoeia generation. Elizalde et al. \cite{related3} utilize a siamese network to jointly map audio and text features into a shared space. However, these methods are restricted by the form of queries. Andreea Maria et al. \cite{related4} exploits Mixture-of-Embedded Experts (MoEE) \cite{MOEE} and Collaborative-Experts (CE) \cite{CE} for the purpose of audio-text retrieval involving natural language descriptions. Lou et al. \cite{ARC} generate audio-text embeddings utilizing pre-trained audio neural networks (PANNs) \cite{panns} and employ NetRVLAD pooling \cite{NetRVLAD}. Mei et al. \cite{ASE} incorporates contrastive learning to increase feature discrimination.
\subsection{Contrastive learning}
Contrastive learning is an unsupervised paradigm designed to enhance the quality of representations by promoting the closeness of positive pairs and the distinctiveness of negative ones. InfoNCE \cite{infoNCE} is proposed to maximize the lower bound of mutual information. NT-Xent \cite{XNENT} aims to promote multi-modal representation learning by building two symmetric contrastive losses.BYOL \cite{BYOL} achieves promising results without negative pairs. Decoupled Contrastive Learning (DCL) \cite{DCL} improves the effectiveness by removing the positive pairs in the denominator of the contrastive loss. Recently, Huang et al. \cite{MACL} proposed Model-Aware Contrastive Learning (MACL) to solve uniformity-tolerance dilemma and gradient reduction. These contrastive methods have been proven effective experimentally. 
\section{Proposed Method}
\subsection{Notations}
In this paper, The multimodal dataset is written as $\mathcal{D} = \{\mathrm{A}, \mathrm{T}\}$, where $\mathrm{A} = [a_1, a_2, ..., a_m]$ and $\mathrm{T} = [t_1, t_2, ..., t_m]$ represent the audio modality and the text modality, respectively. A data batch is denoted as $O = [o_1, o_2, ... , o_b]$, where $b$ is the batch size and $o_j = [a_j, t_j]$ is the $j$th audio-text pair. The relevance between two instances is measured by cosine similarity:
\begin{equation}
	\cos (x, y)=\frac{x^{\mathrm{T}} y}{\|x\|_2\|y\|_2},
\end{equation}
where $x$ and $y$ are two vectors. 
\subsection{Model Formulation}
\begin{figure}[t]
	\centering
	\includegraphics[scale=0.45]{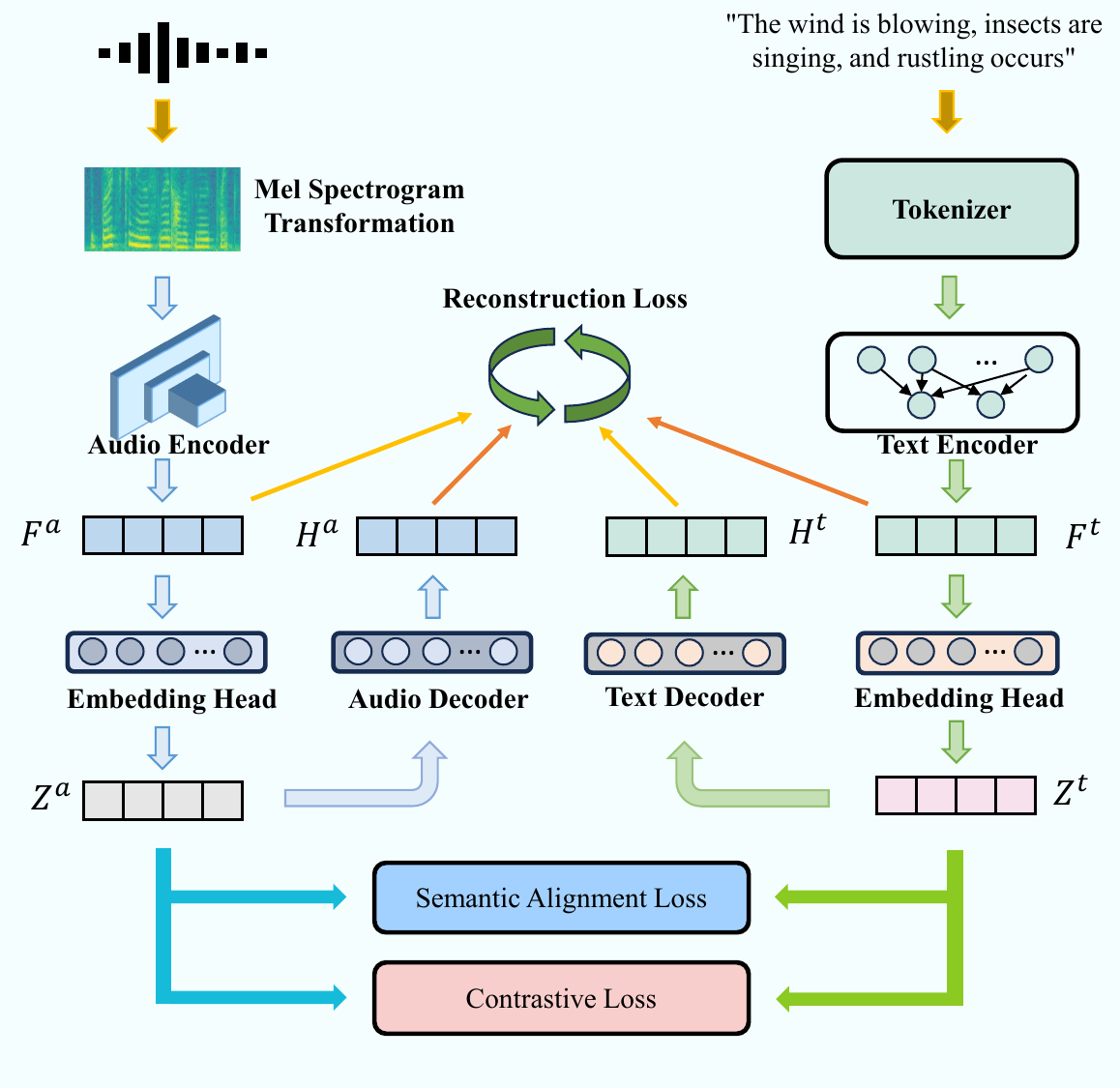}
	\caption{The overall workflow of CLSR.}
	\label{workflow}
\end{figure}

The proposed CLSR framework is depicted in Fig. \ref{workflow}, which comprises three parts: the feature extraction module, the confidence-aware contrastive module and the modality reconstruction module.  
\subsubsection{Feature Extraction Module}
Deep neural networks are prevalent for the strong capacity to capture high-level semantic features. To process audio data, mel frequency spectrogram is widely exploited in speech synthesis (TTS) and voice conversion \cite{cou7,pingan1,pingan2,pingan3}. Mel frequency spectrogram is robust to noise and signal variations in audio signals, for it decomposes the audio signal into several windows, and the signal within each window is stationary, alleviating the disturbance of noise and signal variations. We apply the convolutional network to generate high-level audio features, which is similar to most existing works involving the image modality. To extract semantic features from texts, some literature directly uses bag-of-word vectors and apply MLPs for feature extraction, which loses underlying semantic information. Recently, some researchers propose the textual transformer \cite{bert} to uncover the correlations between words, thus creating more robust semantic-aware features. Inspired by this, we employ the BERT model for textual feature extraction. Therefore, modality-specific features can be obtained:
\begin{equation}
	\begin{aligned}
		&\mathrm{F}^a=\mathrm{E}_a\left(\mathrm{A}, \theta_a\right) \in \mathbb{R}^{b \times d_a}\\
		&\mathrm{F}^t=\mathrm{E}_t\left(\mathrm{T}, \theta_t\right) \in \mathbb{R}^{b \times d_t}
	\end{aligned},
\end{equation}
where $\theta_a$ and $\theta_t$ denote network parameters in two feature extractors, respectively. In this way, semantic correlations can be explored based on the representations.
\subsubsection{Confidence-aware Contrastive Loss}
With the latent representations extracted from the feature extraction modules, we design embedding heads for both modalities to align the modality dimensions. Through the embedding heads, audio and textual features are forced into a shared latent space to facilitate semantic consistency. The extracted features are fed into the embedding head to generate continuous-valued representations. both embedding heads consist of fully connected layers with ReLU activation functions. The process can be described as follows:
\begin{equation}
	\begin{aligned}
		&\mathrm{Z}^a=\mathrm{E}_a\left(\mathrm{F}^a, \epsilon_a\right) \in \mathbb{R}^{b \times c}\\
		&\mathrm{Z}^t=\mathrm{E}_t\left(\mathrm{F}^t, \epsilon_t\right) \in \mathbb{R}^{b \times c}
	\end{aligned},
\end{equation}
where $c$ is the dimension for the shared latent space. 

With the lack of annotated labels, semantic alignment can be achieved by data co-occurrence relationship across modalities. Contrastive learning, as a type of unsupervised learning, aims to learn feature representations that can discriminate between data samples by comparing their similarity. Specifically, for a mini-batch contrastive learning strives to bring data samples of positive pairs closer while simultaneously pushing apart those of negative pairs, and the audio-to-text loss is defined as follows:
\begin{equation}
	\mathcal{L}_{a2t} = -\frac{1}{b}\sum_{i = 1}^{b}\log \frac{\exp \left(a_i t_i^{\mathrm{T}} / \tau\right)}{\exp \left(a_i t_i^{\mathrm{T}} / \tau\right)+\sum_{j=1}^K \exp \left(a_i t_j^{\mathrm{T}} / \tau\right)},
	\label{contrastive}
\end{equation}
where $\{a_i, t_i\}$ is a positive pair, and $K$ is the number of negative pairs for the $i$th instance. And NT-Xent loss \cite{XNENT, ASE}, which is widely used in multi-modal representation learning, can be written as follows: 
\begin{equation}
	\begin{aligned}
		\mathcal{L}_{NT-XENT} & = \mathcal{L}_{a2t} + \mathcal{L}_{t2a}\\
		& = -\frac{1}{b}\sum_{i = 1}^{b}\log \frac{\exp \left(a_i t_i^{\mathrm{T}} / \tau\right)}{\exp \left(a_i t_i^{\mathrm{T}} / \tau\right)+\sum_{j=1}^K \exp \left(a_i t_j^{\mathrm{T}} / \tau\right)}  \\
		& -\frac{1}{b}\sum_{i = 1}^{b}\log \frac{\exp \left(t_i a_i^{\mathrm{T}} / \tau\right)}{\exp \left(t_i a_i^{\mathrm{T}} / \tau\right)+\sum_{j=1}^K \exp \left(t_i a_j^{\mathrm{T}} / \tau\right)}
	\end{aligned}.
	\label{original_con}
\end{equation}
However, this form of contrastive learning is implemented by only taking inter-modal transformation into consideration, neglecting the intra-modal instance separability, which limits the performance of representation learning. CLSR considers expanding Eq. (\ref{original_con}) into the following form:
\begin{equation}
	\mathcal{L}_{con} = \mathcal{L}_{a2t} + \mathcal{L}_{t2a} + \mathcal{L}_{a2a} + \mathcal{L}_{t2t}
	\label{con}
\end{equation}
where $\mathcal{L}_{a2a}$ and $\mathcal{L}_{t2t}$ are contrastive losses applied within the audio modality and the text modality, respectively. Thus, inter-modal and intra-modal contrastive losses are constructed, which further increases multi-modal semantic consistency.

Existing works prove that the temperature parameter is an important factor to undermine hard negative samples, which, however, is often designed empirically and intuitively. Inspired by \cite{MACL}, we propose a confidence-aware temperature scheme to control the temperatures of each sample according to the semantic alignment:
\begin{equation}
	\tau=\tau_0 \cdot \gamma^{\operatorname{Tr}\left(\cos \left(\mathrm{Z}^a, \mathrm{Z}^t\right)\right) / b},
\end{equation}
where $\tau_0$ is the initial temperature, $\gamma > 0$ is the scaling factor, and $\operatorname{Tr}\left(\cos \left(\mathrm{Z}^a, \mathrm{Z}^t\right)\right)$ symbolizes the semantic alignment confidence. Ideally, representations from multi-modal sources are expected to be identical so as to eliminate the heterogeneous gap among modalities, i.e, $\operatorname{Tr}\left(\cos \left(\mathrm{Z}^a, \mathrm{Z}^t\right)\right) / b = 1$. At the beginning of training, a higher temperature is set to impose higher penalties on implicit hard negative samples for better feature consistency. With the increase of iteration, the semantic alignment is improved, and a low temperature is employed to uncover potential positive pairs. 
\begin{algorithm}[t]
	\caption{Optimization for CLSR}\label{optimization}
	\KwIn{Training data $\mathcal{D} = \{\mathbf{A}, \mathbf{T}\}$, code length $c$, batchsize $b$, hyperparameters  \{$\alpha$, $\beta$\}, maximum epoch $T$.}
	\KwOut{Network parameters  $\{\theta_k, \epsilon_k, \gamma_k\}$, $k \in \{a,t\}$ for both the modalities.}
	Initialize the feature extraction module $\theta_a$ and $\theta_t$ with pretrained model, and the rest randomly\;
	\For {$t=1:T$}{
		Randomly sample $O = [o_1, o_2, ... , o_b]$ from $\mathcal{D}$\;
		Extract features $\mathrm{F}^a=\mathrm{E}_a\left(\mathrm{A}, \theta_a\right) $ and $\mathrm{F}^t=\mathrm{E}_t\left(\mathrm{T}, \theta_t\right)$\;
		Encode $\mathrm{Z}^k=\mathrm{E}_k\left(\mathrm{F}^k, \epsilon_k\right)$, $k \in \{a,t\}$\;
		Decode $\mathrm{H}^k=\mathrm{D}_k\left(\mathrm{Z}^k, \gamma_k\right)$, $k \in \{a,t\}$\;
		Compute loss according to E.q. (\ref{overall_func})\;
		Update $\{\theta_k, \epsilon_k, \gamma_k\}$, $k \in \{a,t\}$ with SGD.
	}
	\Return Network parameters  $\{\theta_k, \epsilon_k, \gamma_k\}$, $k \in \{a,t\}$.
\end{algorithm}
\subsubsection{Semantic Consistency Loss}
Cross-modal retrieval aims to bridge the heterogeneous disparity among different modalities, necessitating the establishment of symmetric similarities, i.e., $S_{ij} = S_{ji}$. This correlation can be easily satisfied under the supervised learning setting by leveraging the label information, while it needs to be explicitly constructed in the unsupervised scenario. Thus, the following loss function is designed to maintain semantic consistency:
\begin{equation}
	\mathcal{L}_{sem} = \| cos(Z^a, Z^t) - cos(Z^t, Z^a)\|^2_F,
\end{equation}
where $\|.\|_F$ is the Frobenius norm.

\subsubsection{Reconstruction Loss}
\cite{cou6} has demonstrated that deep feature reconstruction can enhance the cross-modal correlation and reduce the modality gap. Thus, we introduce decoders for both modalities, as illustrated in Fig. \ref{workflow}. Following the extraction of the high-level features, $\mathrm{Z}^a$ and $\mathrm{Z}^t$, we feed these two modality-specific features into the decoders, which can be written as follows: 
\begin{equation}
	\begin{aligned}
		&\mathrm{H}^t=\mathrm{D}_a\left(\mathrm{Z}^a, \gamma_a\right) \in \mathbb{R}^{b \times d_t}\\
		&\mathrm{H}^a=\mathrm{D}_t\left(\mathrm{Z}^t, \gamma_t\right) \in \mathbb{R}^{b \times d_a}
	\end{aligned}.
\end{equation}
Thus, the reconstruction loss of transforming one modality to the other can be measured by:
\begin{equation}
	\mathcal{L}_{rec} = \|F^t - H^t\|_F^2 + \|F^a - H^a\|_F^2.
\end{equation}
Therefore, we can effectively leverage these two compatible attributes and subsequently explore the semantic connections among the information in various modalities to the fullest extent.
\subsubsection{Optimization}
The optimization goal can be written as follows:
\begin{equation}
	\begin{aligned}
		&\min \mathcal{L}= \mathcal{L}_{con} + \alpha \mathcal{L}_{sem} + \beta \mathcal{L}_{rec}\\
		&\text { s.t. } \tau=\tau_0 \cdot \gamma^{\operatorname{Tr}\left(\cos \left(\mathrm{Z}^a, \mathrm{Z}^t\right)\right) / b}
	\end{aligned}.
	\label{overall_func}
\end{equation}
$\alpha$ and $\beta$ are two tunable parameters. The optimization is conducted with SGD. The overall training steps are detailed in Algorithm. \ref{optimization}. 
\section{Experiment}
\begin{table}[t]
	\caption{The R@k results of CLSR and baselines on AudioCaps and Clotho.}
	\begin{tabular}{clclllllll}
		\toprule
		\multicolumn{2}{c}{{\color[HTML]{000000} }}                       & \multicolumn{2}{c}{}                         & \multicolumn{3}{c}{AudioCaps}                                                        & \multicolumn{3}{c}{Clotho}                                                           \\ \cline{5-10} 
		\multicolumn{2}{c}{\multirow{-2}{*}{{\color[HTML]{000000} Task}}} & \multicolumn{2}{c}{\multirow{-2}{*}{Method}} & {\color[HTML]{333333} R@1} & {\color[HTML]{333333} R@5} & {\color[HTML]{333333} R@10} & {\color[HTML]{333333} R@1} & {\color[HTML]{333333} R@5} & {\color[HTML]{333333} R@10} \\ \hline
		\multicolumn{2}{c}{}                                              & \multicolumn{2}{c}{MOEE}                     &            26.6                &                 59.3           &             73.5               &         7.2                     &             22.1              &         33.2                  \\
		\multicolumn{2}{c}{}                                              & \multicolumn{2}{c}{CE}                       &       27.6                      &             60.5                  &           74.7                    &      7.0                         &         22.7                &      34.6                   \\
		\multicolumn{2}{c}{}                                              & \multicolumn{2}{c}{ARC}                      &               33.3            &            65.3                &      80.6                      &        13.0                    &                30.5            &          45.4                  \\
		\multicolumn{2}{c}{}                                              & \multicolumn{2}{c}{ASE}                      &      38.8                       &             71.5               &            83.1                 &        13.4                      &             36.1               &            49.3                \\
		\multicolumn{2}{c}{\multirow{-5}{*}{A2T}}                         & \multicolumn{2}{c}{CLSR}                     &   \textbf{42.2}                        &        \textbf{73.3}                    &         \textbf{84.5}                    &        \textbf{15.3}                    &          \textbf{36.3}                  &        \textbf{49.6}                    \\ \hline
		\multicolumn{2}{c}{}                                              & \multicolumn{2}{c}{MOEE}         & 23.0                   &            55.7                &           71.0  &                 6.0                              &                20.8            &                   32.3         \\
		\multicolumn{2}{c}{}                                              & \multicolumn{2}{c}{CE}                       &      23.6                       &            56.2                &                71.4            &            6.7                 &               21.6              &             33.2               \\
		\multicolumn{2}{c}{}                                              & \multicolumn{2}{c}{ARC}                      &             29.3               &               60.2             &               79.3             &       13.1                     &            28.2                &             45.1               \\
		\multicolumn{2}{c}{}                                              & \multicolumn{2}{c}{ASE}                      &      33.4                        &              69.1              &             81.7               &            13.2                 &              35.8              &             49.6               \\
		\multicolumn{2}{c}{\multirow{-5}{*}{T2A}}                         & \multicolumn{2}{c}{CLSR}                     &      \textbf{34.1}                      &               \textbf{70.0}             &             \textbf{83.7}               &              \textbf{13.4}              &             \textbf{36.2}               &             \textbf{50.3}               \\ 		\toprule
	\end{tabular}
	\label{RK}
\end{table}
\subsection{Datasets}
The effectiveness of CLSR is verified on two standard datasets, i.e., AudioCaps \cite{Audiocaps} and Clotho \cite{Clotho}.

\textbf{AudioCaps} comprises approximately 50,000 audio clips, each lasting for 10 seconds, sourced from AudioSet \cite{audioset}. Among these, 49,274 audio clips have been chosen for the training set, each accompanied by its respective textual description. Furthermore, the validation set comprises 494 audio clips, while the test set contains 957 audio clips. Each audio clip in both sets is paired with five corresponding captions.

\textbf{Clotho} contains audio clips ranging from 15 to 30 seconds, annotated by corresponding textual data. The training set comprises 3,839 audio clips, while both the validation and test sets include 1,045 audio clips each. Additionally, each audio clip is paired with five corresponding textual descriptions. 
\subsection{Experimental Setups}
Following \cite{ASE}, the log Mel-spectrogram is computed using a Hanning window with 1024 points and a hop size of 320 points, resulting in 64 mel bins. The maximum training epoch is set to be 50, with Adam as the optimizer. The learning rate is set as $1e^{-4}$ with a decay rate of $1/10$ per 20 epochs. The initial temperature $\tau_0$ is equal to $0.07$, the scaling factor $\gamma$ is set to 1.2, and the dimension of the output embedding is $1024$. $\alpha$ and $\beta$ are set to be 1 and 0.1, respectively. Experiments are carried out using a single NVIDIA Tesla V100 GPU.

The evaluation metric employed is Recall at rank $k$ (R@k), which indicates whether the most relevant content to a query appears at the top rank $k$. The R@k metric is scaled between 0 and 1, and higher values signify better performance, where the top-ranked retrieval result is more likely to be related to the query.
\subsection{Evaluation}

We compare CLSR with several baselines, i.e., MOEE \cite{MOEE}, CE \cite{CE}, ARC \cite{ARC}, and ASE \cite{ASE}. It should be noticed that MOEE and CE are two frameworks used for text-video retrieval, and, following \cite{overview}, we adapt it for text-audio retrieval. We evaluate CLSR and these baselines on AudioCaps and Clotho, and the experimental results including R@1, R@5 and R@10, are detailed in Table \ref{RK}. Based on
the experimental results, it can be observed that:
\begin{itemize}
	\item CLSR consistently achieves superior R@k results in both Audio-to-Text and Text-to-Audio tasks across both datasets. CLSR considers enlarging the distance of negative pairs to enhance instance discrimination, contributing to more robust latent embedding. For contrastive learning, to alleviate the limitation of a small batch size, a reweighting strategy is imposed to adaptively adjust weight of different positive pairs. 
	\item In most cases, R@k results on Audio-to-Text task  tend to surpass those for the Text-to-Audio task. This trend may be attributed to the rich semantic information in textual data compared to audio data.
\end{itemize}
\subsection{Ablation Study}
For CLSR, multi-modal contrastive learning, semantic consistency and the reconstruction loss are adopted for CMR representation learning. In this subsection, ablation study is conducted to analyze their effects. From CLSR, four variants are derived, i.e., CLSR-s, CLSR-t, CLSR-k and CLSR-m. CLSR-s removes the intra-modal contrastive loss, i.e., the third term and the fourth term in Eq. (\ref{con}). CLSR-t drops the confidence-aware temperature control strategy. CLSR-k adopts no semantic consistency loss, which is the second term in E.q (\ref{overall_func}). CLSR-m discards the reconstruction loss. CLSR's performance is assessed in comparison to these four variants on AudioCaps.
\begin{table}[t]
	\caption{The Experiment Results of CLSR with four variants on AudioCaps.}
	\begin{tabular}{clclllllll}
		\toprule
		\multicolumn{2}{c}{{\color[HTML]{000000} }}                       & \multicolumn{2}{c}{}                         & \multicolumn{3}{c}{AudioCaps}                                                        & \multicolumn{3}{c}{Clotho}                                                           \\ \cline{5-10} 
		\multicolumn{2}{c}{\multirow{-2}{*}{{\color[HTML]{000000} Task}}} & \multicolumn{2}{c}{\multirow{-2}{*}{Method}} & {\color[HTML]{333333} R@1} & {\color[HTML]{333333} R@5} & {\color[HTML]{333333} R@10} & {\color[HTML]{333333} R@1} & {\color[HTML]{333333} R@5} & {\color[HTML]{333333} R@10} \\ \hline
		\multicolumn{2}{c}{}                                              & \multicolumn{2}{c}{CLSR-s}                     &          39.8                 &         71.5                   &           83.1             &            13.7                  &              35.9            &             49.1             \\
		\multicolumn{2}{c}{}                                              & \multicolumn{2}{c}{CLSR-t}                       &        41.5                 &          71.3                 &          82.7                  &            14.8               &      35.7                &         48.9            \\
		\multicolumn{2}{c}{}                                              & \multicolumn{2}{c}{CLSR-k}                      &        41.9               &          71.7                 &                83.4          &     14.6               &          36.1            &         47.1                                                              \\
		\multicolumn{2}{c}{}                                              & \multicolumn{2}{c}{CLSR-m}                      &         41.8                 &         72.6               &           84.2               &          14.7                &     35.7                    &          48.0                \\
		\multicolumn{2}{c}{\multirow{-5}{*}{A2T}}                         & \multicolumn{2}{c}{CLSR}                     &   \textbf{42.2}                        &        \textbf{73.3}                    &         \textbf{84.5}                    &        \textbf{15.3}                    &          \textbf{36.3}                  &        \textbf{49.6}                    \\ \hline
		\multicolumn{2}{c}{}                                              & \multicolumn{2}{c}{CLSR-s}         &            33.5      &              69.5            &    81.7        &              13.3                                &            35.9           &              48.4            \\
		\multicolumn{2}{c}{}                                              & \multicolumn{2}{c}{CLSR-t}                       &        33.9                   &           68.9              &          81.6                 &            13.2               &               36.1             &      48.6                \\
		\multicolumn{2}{c}{}                                              & \multicolumn{2}{c}{CLSR-k}                      &         33.6                  &            69.2             &        81.9               &         13.1                  &        35.9                &          48.6                \\
		\multicolumn{2}{c}{}                                              & \multicolumn{2}{c}{CLSR-m}                      &       33.8                  &             69.7             &             82.3           &            13.2              &         35.8              &           48.0               \\
		\multicolumn{2}{c}{\multirow{-5}{*}{T2A}}                         & \multicolumn{2}{c}{CLSR}                     &      \textbf{34.1}                      &               \textbf{70.0}             &             \textbf{83.7}               &              \textbf{13.4}              &             \textbf{36.2}               &             \textbf{50.3}               \\ 		\toprule
	\end{tabular}
	\label{ablation}
\end{table}

The R@k scores are listed in TABLE \ref{ablation}. It can be shown that CLSR generally surpasses these variants. Concretely, CLSR-s achieves low R@k results, which demonstrates the effectiveness of the expanded intra-modal contrastive loss for learning robust latent representations. CLSR-t, CLSR-k achieve better results than CLSR-s, which are still inferior to CLSR. This demonstrate that the adaptive temperature scheme and semantic consistency contribute to improving the retrieval results. The performance of CLSR-m is satisfying, but still lower than CLSR, validating the efficacy of the reconstruction loss.
\section{Conclusion}
In this paper, a novel CMR method, dubbed as Contrastive Latent Space Reconstruction (CLSR), is introduced to address the audio-text retrieval. CLSR is implemented in an unsupervised fashion, employing contrastive learning to dynamically increase the separation between negative pairs while reducing the gap between positive pairs. We expand the multi-modal contrastive loss by taking intra-modal separability into consideration, which improves the robustness of the learned representations. Besides, the temperature setting is adaptive along with semantic alignment, allowing CLSR to excavate latent correlations between samples. CLSR utilizes semantic consistency and modality reconstruction to increase the discrimination of sample features. 
Experimental results across two datasets illustrate that CLSR surpasses certain state-of-the-art cross-modal methods.
\section{Acknowledgement}
This paper is supported by the Key Research and Development Program of Guangdong Province under grant No.2021B0101400003. Corresponding author is Jianzong Wang from Ping An Technology (Shenzhen) Co., Ltd (jzwang@188.com).
\bibliographystyle{IEEEtran.bst}
\bibliography{mybib.bib}

\end{document}